\documentclass[aps,prl,twocolumn,superscriptaddress,showpacs]{revtex4-1}
\usepackage{amsmath,amssymb,amsfonts,amsthm}
\usepackage{graphicx}
\usepackage{natbib}
\usepackage{dcolumn}   % needed for some tables
\usepackage{epstopdf}
\usepackage[caption=false]{subfig}

\begin{document}

 \newcommand{\breite}{1.0} %  for twocolumn

\newtheorem{prop}{Proposition}
\newtheorem{cor}{Corollary}

\newcommand{\be}{\begin{equation}}
\newcommand{\ee}{\end{equation}}

\newcommand{\bea}{\begin{eqnarray}}
\newcommand{\eea}{\end{eqnarray}}

\newcommand{\Reals}{\mathbb{R}}     % Reals
\newcommand{\Com}{\mathbb{C}}       % Complex #
\newcommand{\Nat}{\mathbb{N}}       % Natural #
\newcommand{\ignore}[1]{}

\newcommand{\id}{\mathbbm{1}}    

\newcommand{\Real}{\mathop{\mathrm{Re}}}
\newcommand{\Imag}{\mathop{\mathrm{Im}}}

\def\O{\mbox{$\mathcal{O}$}}   % Order epsilon ... 
\def\F{\mathcal{F}}			% FourierTrafo
\def\sgn{\text{sgn}}

\newcommand{\deo}{\ensuremath{\Delta_0}}
\newcommand{\dea}{\ensuremath{\Delta}}
\newcommand{\ak}{\ensuremath{a_k}}
\newcommand{\ad}{\ensuremath{a^{\dagger}_{-k}}}
\newcommand{\sx}{\ensuremath{\sigma_x}}
\newcommand{\sz}{\ensuremath{\sigma_z}}
\newcommand{\spl}{\ensuremath{\sigma_{+}}}
\newcommand{\smi}{\ensuremath{\sigma_{-}}}
\newcommand{\alk}{\ensuremath{\alpha_{k}}}
\newcommand{\bk}{\ensuremath{\beta_{k}}}
\newcommand{\ok}{\ensuremath{\omega_{k}}}
\newcommand{\vd}{\ensuremath{V^{\dagger}_1}}
\newcommand{\vi}{\ensuremath{V_1}}
\newcommand{\vo}{\ensuremath{V_o}}
\newcommand{\zc}{\ensuremath{\frac{E_z}{E}}}
\newcommand{\xc}{\ensuremath{\frac{\Delta}{E}}}
\newcommand{\xd}{\ensuremath{X^{\dagger}}}
\newcommand{\aok}{\ensuremath{\frac{\alk}{\ok}}}
\newcommand{\tpw}{\ensuremath{e^{i \ok s }}}
\newcommand{\tpe}{\ensuremath{e^{2iE s }}}
\newcommand{\tmw}{\ensuremath{e^{-i \ok s }}}
\newcommand{\tme}{\ensuremath{e^{-2iE s }}}
\newcommand{\epls}{\ensuremath{e^{F(s)}}}
\newcommand{\emis}{\ensuremath{e^{-F(s)}}}
\newcommand{\epl}{\ensuremath{e^{F(0)}}}
\newcommand{\emi}{\ensuremath{e^{F(0)}}}

\newcommand{\lr}[1]{\left( #1 \right)}
\newcommand{\lrs}[1]{\left( #1 \right)^2}
\newcommand{\lrb}[1]{\left< #1\right>}
\newcommand{\nbt}{\ensuremath{\lr{ \lr{n_k + 1} \tmw + n_k \tpw  }}}

\newcommand{\om}{\ensuremath{\omega}}
\newcommand{\dw}{\ensuremath{\Delta_0}}
\newcommand{\wbp}{\ensuremath{\omega_0}}
\newcommand{\dv}{\ensuremath{\Delta_0}}
\newcommand{\vbp}{\ensuremath{\nu_0}}
\newcommand{\vplus}{\ensuremath{\nu_{+}}}
\newcommand{\vminus}{\ensuremath{\nu_{-}}}
\newcommand{\wplus}{\ensuremath{\omega_{+}}}
\newcommand{\wminus}{\ensuremath{\omega_{-}}}
\newcommand{\uv}[1]{\ensuremath{\mathbf{\hat{#1}}}} % for unit vector
\newcommand{\abs}[1]{\left| #1 \right|} % for absolute value
\newcommand{\avg}[1]{\left< #1 \right>} % for average
\let\underdot=\d % rename builtin command \d{} to \underdot{}
\renewcommand{\d}[2]{\frac{d #1}{d #2}} % for derivatives
\newcommand{\dd}[2]{\frac{d^2 #1}{d #2^2}} % for double derivatives
\newcommand{\pd}[2]{\frac{\partial #1}{\partial #2}} 
% for partial derivatives
\newcommand{\pdd}[2]{\frac{\partial^2 #1}{\partial #2^2}} 
% for double partial derivatives
\newcommand{\pdc}[3]{\left( \frac{\partial #1}{\partial #2}
 \right)_{#3}} % for thermodynamic partial derivatives
\newcommand{\ket}[1]{\left| #1 \right>} % for Dirac bras
\newcommand{\bra}[1]{\left< #1 \right|} % for Dirac kets
\newcommand{\braket}[2]{\left< #1 \vphantom{#2} \right|
 \left. #2 \vphantom{#1} \right>} % for Dirac brackets
\newcommand{\matrixel}[3]{\left< #1 \vphantom{#2#3} \right|
 #2 \left| #3 \vphantom{#1#2} \right>} % for Dirac matrix elements
\newcommand{\grad}[1]{\gv{\nabla} #1} % for gradient
\let\divsymb=\div % rename builtin command \div to \divsymb
\renewcommand{\div}[1]{\gv{\nabla} \cdot #1} % for divergence
\newcommand{\curl}[1]{\gv{\nabla} \times #1} % for curl
\let\baraccent=\= % rename builtin command \= to \baraccent
%%%%%%%%%%%%%%%%%%%%%%%%%%%%%%%%%%%%%%%%%%%%%
% End Definitions
%%%%%%%%%%%%%%%%%%%%%%%%%%%%%%%%%%%%%%%%%%%%%

%Title of paper
\title{Chiral Prethermalization in supersonically split condensates}

\author{Kartiek Agarwal}
\affiliation{Physics Department, Harvard University, Cambridge, Massachusetts 02138, USA}
\email[]{agarwal@physics.harvard.edu}
\author{Emanuele G. Dalla Torre}
\affiliation{Physics Department, Harvard University, Cambridge, Massachusetts 02138, USA}
\author{Bernhard Rauer}
\affiliation{Vienna Center for Quantum Science and Technology, Atominstitut, TU Wien, Stadionallee 2, 1020 Vienna, Austria}
\author{Tim Langen}
\affiliation{Vienna Center for Quantum Science and Technology, Atominstitut, TU Wien, Stadionallee 2, 1020 Vienna, Austria}
\author{J\"{o}rg Schmiedmayer}
\affiliation{Vienna Center for Quantum Science and Technology, Atominstitut, TU Wien, Stadionallee 2, 1020 Vienna, Austria}
\author{Eugene Demler}
\affiliation{Physics Department, Harvard University, Cambridge, Massachusetts 02138, USA}

\date{\today}
\begin{abstract}
We study the dynamics of phase relaxation between a pair of one-dimensional condensates created by a supersonic unzipping of a single condensate. We use the Lorentz invariance of the low energy sector of such systems to show that dephasing results in an unusual prethermal state, in which right- and left-moving excitations have different, Doppler-shifted temperatures. The chirality of these modes can be probed experimentally by measuring the interference fringe contrasts with the release point of the split condensates moving at another supersonic velocity. Further, an accelerated motion of the release point can be used to observe a space-like analogue of the Unruh effect. A concrete experimental realization of the quantum zipper for a BEC of trapped atoms on a atom chip is outlined. We interpret these results in the context of the general question of the Lorentz transformation of temperature, and the close analogy with the dipolar anisotropy of the Cosmic Microwave Background. 
\end{abstract}
\maketitle

\paragraph*{Introduction} Trapped gases of ultra-cold atoms today provide the most remarkable examples of nearly isolated quantum systems. While traditional condensed matter systems are typically difficult to isolate from external noise, these artificial systems have been successfully engineered to be sufficiently decoupled \cite{Greiner,Sidorov} from the environment, so that one can assume a unitary evolution of the system over long time scales. These developments, have in turn, reinvigorated theoretical interest in the study of out-of-equilibrium dynamics of isolated quantum systems. 
 
  Amidst such interest, one-dimensional (1D) systems have garnered particular attention because the non-equilibrium behavior in these systems is enhanced \cite{Kinoshita06} due to limited phase space available for scattering and equilibration. Further, in many cases, these systems exhibit an emergent Lorentz invariance (and sometimes a larger conformal symmetry) - for example, while the constituent atoms of 1D gases obey only Galilean invariance, the collective modes, described by the Luttinger Liquid theory, exhibit a richer Lorentz invariance. These symmetries make calculations feasible, and lead to many universal features in non-equilibrium dynamics, such as scaling laws for the growth of domains \cite{Lamacraft,Polkovnikov}, light-cone spreading of correlations \cite{Cheneau,Langen} and the relaxation of observables \cite{Calabrese}. Such features have also been seen in experiments \cite{Sadler,Chen,Ronzheimer,Trotzky}. 

 One of the basic tools for studying non-equilibrium dynamics is the `quantum quench'. The protocol involves preparing the system in the ground state of a Hamiltonian $H_0$, and, subsequently evolving it with another Hamiltonian $H$, for which the initial state is not an eigenstate. Typical scenarios for quenches include a sudden turning off of the confining potential of the 1D gas \cite{Meisner,Rigol2,Rigol3}, the interaction between the constituent particles \cite{Shashi}, or, in a more abstract case, the excitation gap of the low energy excitations of the system \cite{Cazalilla,Dalla}. Importantly, most previous studies have focussed on $sudden$ (and spatially homogeneous) quenches, that is, the transition from $H_0$ to $H$ happens over a time-scale shorter than any other time-scale in the problem. The study of such quenches, however, does not fully utilize the rich symmetry of the low energy physics of these systems. 
 \begin{figure}
\includegraphics[width = 2.8in]{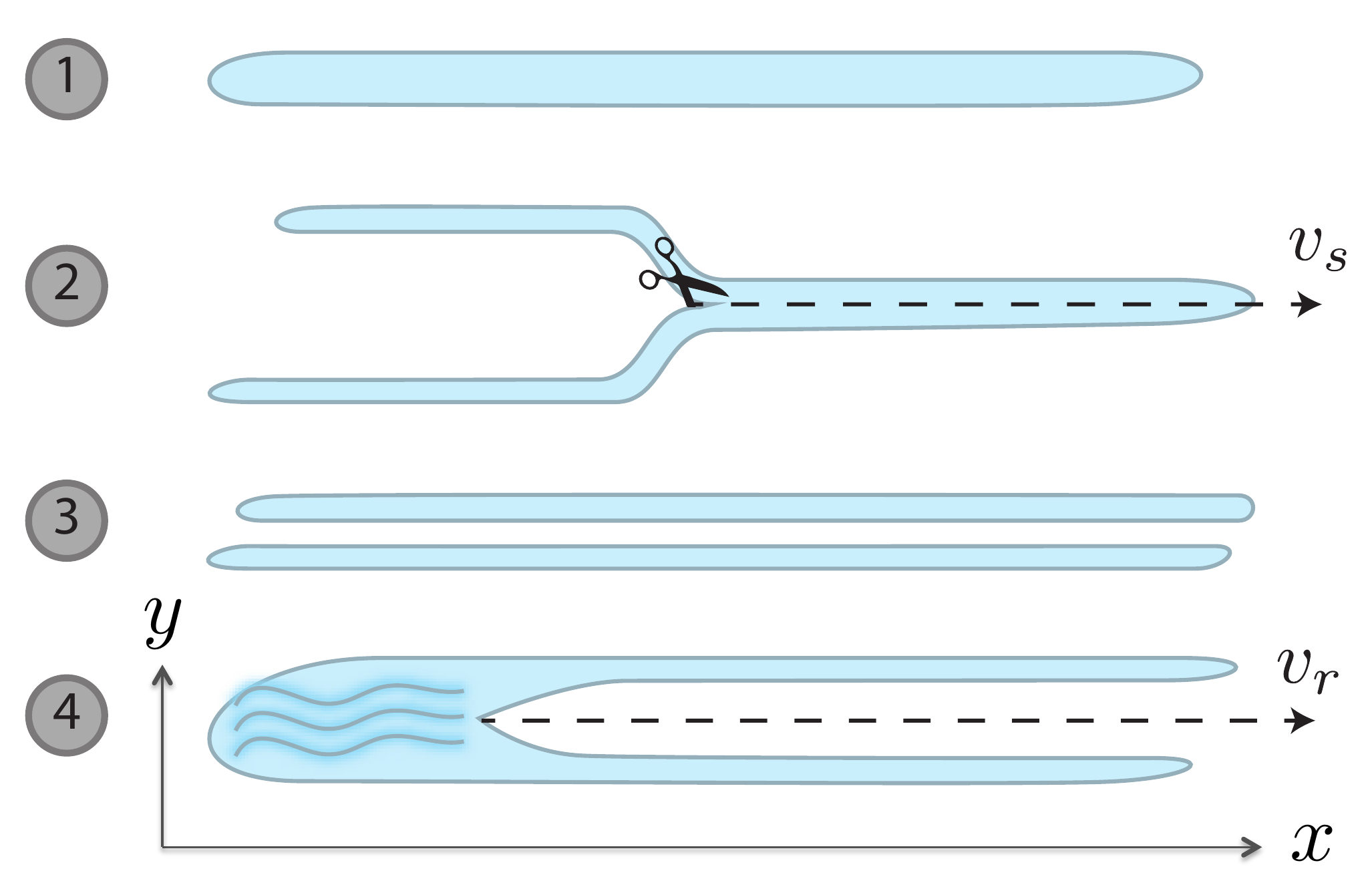}
\caption{The experimental protocol : 1) Preparation of a 1D condensate 2) The condensate is split by a perturbation traveling at a superluminal velocity $v_s > c$ 3) The split condensates are allowed to freely evolve in time 4) They are released at a superluminal velocity $\abs{v_r} > c$ and the interference fringes are recorded. \vspace{-0.65 cm}}
%\caption{The experimental protocol}
\label{fig:protocol}
\end{figure} 

 The aim of this Letter is to describe how a new class of quantum quenches, wherein the Hamiltonian is continually perturbed or `quenched' along a relativistic, supersonic trajectory, can be analyzed by using this Lorentz symmetry to full effect. In particular, due to the supersonic character of the quench, and the Lorentz invariance of ground state correlations, we show that such class of quenches can be mapped to the usual sudden quench case via a Lorentz boost.  To motivate the analysis of such a quench, we describe our theoretical problem in the context of recent experiments \cite{Gring}  studying the relaxation of the phase difference between the halves of a coherently split quasi-1D condensate. After the splitting, the phase difference evolves freely and can be described by a Luttinger Hamiltonian \cite{Takuya}, with an initial state that appropriately represents the squeezing of the relative phase to zero (bar quantum fluctuations) at the time of the quench. We propose a modification of the sudden quench protocol to one where the splitting happens along a `knife-edge' that travels through the condensate at a supersonic velocity $v_s$ (see Fig. \ref{fig:protocol}). Such a protocol may be realized by using an inhomogeneous set of RF fields that create a propagating splitting potential (see SI \ignore{\ref{sec:protocolexp}})

An interesting prediction of previous studies \cite{Bistritzer,Takuya} describing the sudden quench was that the system eventually evolves to a state with thermal-looking correlations at a temperature $T_0 = \mu/2$, where $\mu$ is the chemical potential of each half condensate. Such predictions were confirmed by experiments \cite{Gring}. Now, since our problem can be mapped to the sudden quench problem in a Lorentz boosted frame, one can interpret our proposed experiment as one that creates a $moving$ relativistic body, which is at a temperature $T_0$ in its rest frame. Therefore, the analysis of our problem also sheds light on a much debated problem in relativistic thermodynamics - what is the temperature of a moving body, or equivalently, how does temperature transform under a Lorentz transformation \cite{Einstein,Planck,Ott,Kibble,Landsberg,Callen,Sewell,Landsberg2}? To this end, we find that while correlations of our time-dependent problem approach a prethermal state, they display a chiral asymmetry - left-moving excitations appear cooler, at a temperature $T_0 / \eta_s$, while right-moving ones appear hotter, by the same Doppler shift $\eta_s = \sqrt{\frac{1+c/v_s}{1-c/v_s}}$. However, equal-time correlations of the system are described by an average of these two temperatures, $T = \left( \eta_s+ 1/\eta_s \right) T_0/2$. To bring out the chirality, we propose probing unequal-time correlations \cite{Essler} of the phase. In particular, releasing the condensate at a supersonic velocity $v_r$ (see Fig. \ref{fig:protocol}) gives us access to correlations of the phase field at space-time points obeying the relation $t = x/v_r$. We find that such correlations (characterized by different $v_r$) exhibit the whole range of effective temperatures $T_0/ \eta_s < T < \eta_s T_0$. The fact that different correlations can be characterized by different temperatures clearly suggests that a definitive answer to the relativistic transformation law for temperature cannot be given (see \cite{Callen} and references therein).
\paragraph*{Model}
To describe the splitting process, we consider the condensate as a system of two inter-coupled 1D Luttinger liquids, whose mutual coupling is destroyed in the process of splitting. The Hamiltonian that describes the dynamics of the phase difference field $\phi$ is
\be
H (t) = \int dx \; \left( \frac{\rho}{4 m} \left( \partial_x \phi \right)^2 + g n^2 + J \Theta (x - v_s t ) \phi^2 \right) .
\label{eq:Ht}
\ee
Here $n$ is the number fluctuations conjugate to the phase $\phi$, $\rho$ is the density of each half of the condensate, $g$ characterizes the strength of contact interactions, $v_s$ is the velocity at which the splitting perturbation travels down the condensate, and $\Theta$ is the Heaviside function. 
The value of the coupling $J$ is set by the process of splitting. We choose $J = g \rho^2$ so that the initial correlations (in the ground state of H(-$\infty$)) match the correlation one expects to see at the time of splitting (gaussian correlations with moment $\avg{n(x)n(x')} = \rho/2 \delta(x-x')$, see Bistritzer et al. \cite{Bistritzer}). Since $v_s$ is supersonic, light-cone physics guarantees that correlations of $H(-\infty)$ persist at every point until the time of splitting. Thus, $H(t)$ captures the two essential features of our problem - (i) it correctly describes the correlations at the time of splitting and, (ii) the dynamics after the splitting is governed by a Luttinger model. 
%The value of the coupling $J$ is set by the process of splitting. To determine it, we follow the arguments developed by Bistritzer et al. \cite{Bistritzer}, who analyzed the problem of the sudden spatially uniform splitting of a condensate. They argue that when the condensate is split, particles enter into a superposition, occupying either half of the split condensate with equal amplitude. In the thermodynamic limit, this implies that $n$ must be a gaussian distributed field, with uncertainty $\sqrt{\rho/2}$ (or correlation $\avg{n(x)n(x')} = \rho/2 \delta(x-x')$). For $H(t)$ to correctly describe our problem, we require these correlations to exist at every point $x$ at the time of splitting $t = x/v_s$. To achieve this, we choose $J = g \rho^2$ (see Supplemetary Materials \ref{sec:diag}) such that the ground state of the Hamiltonian in Eq. (\ref{eq:Ht}) at $t = -\infty$ has precisely these correlations. But to ensure that these correlations persist (under the evolution by $H(t)$) at every point $x$ at the time of splitting, we require that the splitting velocity $v_s$ is greater than the speed of sound of the excitations of $H$, that is, $v_s > c = \sqrt{ \rho g /m}$. Due to this constraint, we limit ourselves to evaluating the case of superluminal splitting. 
%\footnote{Strictly speaking, there is no Lorentz symmetry for a finite system. However, as long as we work in the thermodynamic limit, with a supersonic splitting velocity, this approach remains valid. For a detailed discussion, see SI \ignore{\ref{sec:fin}}}.
\paragraph*{Dynamics} In order to simplify calculations, we analyze the problem in a frame that is Lorentz-boosted from the lab frame by velocity $u_s = (c/v_s) c < c$. The Lorentz transformation $\mathcal{L} : \{ x_i , t_i\} \rightarrow \{ x'_i, t'_i\}$ with $x' = \gamma_s (x-u_s t)$, $t' = \gamma_s (t - u_s x / c^2)$, and $\gamma_s = \frac{1}{\sqrt{1-(u_s/c)^2}} = \frac{1}{\sqrt{1-(c/v_s)^2}}$, that achieves this boost, results in a convenient form of the splitting perturbation $\Theta_s (x - v_s t) \rightarrow \Theta_s(-t')$ , which appears to be spatially uniform and sudden in this frame \cite{Guerreiro}. The dynamics of the system are then governed by the following well-studied quench \cite{Cazalilla}
\begin{align}
H'(t') &= \int dx' \; \left( \frac{\rho}{4 m} \phi'^2 + g n'^2 + J  \Theta(-t') \phi'^2 \right),
\label{eq:Hb}
\end{align}
where the field $\phi'$ satisfies the relation $\phi'(x'_i,t'_i) = \phi(x_i,t_i)$, and $n'$ is conjugate to $\phi'$.  
In general, such a coordinate transformation may simplify the dynamics, but complicates the form of initial correlations. In our case, prior to any splitting (for all times $t < -\infty$), the system resides in the ground state of a Lorentz invariant system described by the Hamiltonian $H(t < -\infty)$. Now, all vacuum correlation of a scalar field governed by a Lorentz invariant action are invariant under Lorentz transformations \cite{Peskin}. Mathematically, this means $\avg{ \phi(x_1,t) \phi(x_2,t)} = \avg{\phi( \mathcal{L}(x_1,t)) \phi(\mathcal{L}(x_2,t)} = \avg{ \phi'(x_1,t) \phi'(x_2,t)}$. Therefore, the initial correlations of the field $\phi'$ in the boosted frame are of the same form as those of the field $\phi$ in the laboratory frame. Moreover, these correlations correspond to the ground state of the Hamiltonian $H(-\infty)$ which is precisely of the same form as $H'(t'<0)$. Therefore, in this boosted frame, the system starts in the ground state of $H'(t'<0)$. At precisely $t' = 0$, which is the trajectory of the splitting perturbation in the boosted frame, the system is perturbed, and subsequently evolves with $H'(t'>0)$. Formally, the problem in this boosted frame is identical to that of the sudden quench case considered by previous authors. 
\paragraph*{Distribution Functions} To characterize the dynamics, we calculate the distribution function (DF) $P(\alpha,t)$ of the spatially integrated phase contrast $\alpha = \abs{\int^{l/2}_{-l/2} \frac{dx}{l} e^{i \phi(x,t)}}^2$. Since we are interested in unequal time correlations of our system, the phase $\phi(x,t)$ in $\alpha$, will correspond to times $t$ related to the position $x$ as $t = t_0 + x/v_r$, with $\abs{v_r} > c$. 
%\be
%\avg{\abs{\int^{l/2}_{-l/2} dx \; e^{i \phi(x,t)} / l}^{2m}} = \int d \alpha P(\alpha,t_0,v_s,v_r) \alpha^m .
%\label{eq:pdef}
%\ee

The distribution functions for equal time correlations were measured in experiments by repeatedly recording the value of the integrated phase contrast over many experimental runs \cite{PolkovnikovPNAS,Kuhnert,Gring,Langen}. Theoretically, the distributions characterizing the correlations of the field $e^{i \phi'(x'_i,t'_i)}$ at equal times $t'_i$ were evaluated by Kitagawa $et$ $al.$ \cite{Takuya} in their analysis of the sudden splitting case. This approach can be directly generalized to evaluate unequal-time correlations as long as the points $\{(x'_i, t'_i)\}$ (or equivalently $\{(x_i, t_i)\}$ in lab coordinates) are space-like separated. Mathematically, this ensures that field operators at such points commute, and makes calculations feasible. Physically, this is a constraint, because experiments cannot record correlations of operators at time-like separated points, without altering the system during the measurement process. The condition $\abs{v_r} > c$ precisely ensures that $\{(x_i , t_i )\}$, are space-like separated. Once correlations are calculated for the field $\phi'$, we simply re-express the results in terms of the relevant lab frame coordinates, to get the result for the distribution functions $P(\alpha, l,t_0, v_s, v_r)$. Following the procedure of \cite{Takuya}, we arrive at the result (see SI \ignore{\ref{sec:solve}})
\begin{align}
P &= \prod_k \int \frac{d\theta_k}{2 \pi} \frac{dr^2_k}{2} e^{- r^2_k/2} \delta \left( \alpha - \abs{\int^{l/2}_{-l/2} \frac{dx}{l} e^{i \chi_d}}^2\right), \label{eq:pdist} \\
\chi_d &= \sum_{\epsilon = \pm 1} \int \frac{dk}{\sqrt{2 \pi}} \sqrt{\frac{g}{c \abs{k}}} A(k,\epsilon) r_k \sin{( (\eta_s)^{\epsilon} k (x - \epsilon c t)+ \theta_k)}, \nonumber
\end{align}
where $\eta_s (v_s) = \sqrt{\frac{1 + u_s/c}{1 - u_s/c}} = \sqrt{\frac{1 + c/v_s}{1 - c/v_s}} > 1$ is the relativistic Doppler shift, associated with the Lorentz boost at velocity $u_s$, and $A(k,\epsilon) = \frac{\sgn{(k)}}{2} \left( \sqrt{\frac{2 g \rho}{c \abs{k}}} + \sgn{(k)} \epsilon \sqrt{\frac{c \abs{k}}{2 g \rho}}  \right)$. This result has a clear physical interpretation - the splitting process generates waves of momentum $k$ with a magnitude proportional to $r_k$ and a random phase $\theta_k$. Each configuration $\{r_k,\theta_k\}$ serves as an initial condition for which we can predict the exact evolution of the phase field $\phi(x,t)$, which shows up as the real function $\chi_d(x,t)$. The probability that the integrated phase contrast takes a given value $\alpha$ is found by performing an integral over the phase space of all initial conditions $\{r_k,\theta_k\}$ with the appropriate statistical weight. The factors of $\eta_s$ accompanying right-moving waves $\propto (x-ct)$ and $1/\eta_s$ accompanying left-moving waves $\propto (x+ct)$ simply indicate the relativistic blue and red shifting of the corresponding sets of waves. However, it is important to note that when we measure unequal time correlations (characterized by finite $v_r$), we will set $t(t_0,x) = t_0 + x/v_r$ which will in general modify these dilation factors to $a_R = \eta_s (1 - c/v_r)$ and $a_L = (1 + c/v_r)/\eta_s$. These factors will figure prominently when we see how these DFs behave in the long time limit $t_0 \rightarrow \infty$. 
\paragraph*{Prethermalization} The tendency of an integrable system to flow into a state with correlations describable by a Generalized Gibbs Ensemble (GGE) is called `prethermalization'.  To explore the process of prethermalization in our problem, we compare the DFs of the integrated phase contrast obtained for our dynamical system to that obtained for two independent, thermal condensates. The DF for two thermal condensates at a temperature $T$ can be obtained using an entirely similar approach to the one used above for our dynamical problem. The thermal result is simply a modification of the results of Eq. (\ref{eq:pdist}) with $\chi_d$ being replaced by $\chi_T$ given by \cite{Imambekov}
\begin{align}
\chi_T (x, r_k, \theta_k) &= \int \frac{dk}{\sqrt{2 \pi}} r_k \sqrt{f_T (k)} \sin{(kx + \theta_k)}, \label{eq:chit} \\
f_T (k) &= \frac{g}{c \abs{k}} \coth{\left(\frac{c\abs{k}}{2 k_B T}\right) } \approx \frac{g}{c \abs{k}} \frac{ 2 k_B T}{c \abs{k}}. \label{eq:ftk}
\end{align}
Here, we note that in the classical limit of $T$ being large, the amplitude of each of the waves, $f_T (k)$ is $\propto T/k^2$. This is in accordance with the classical equipartition theorem, which guarantees that each mode in our quadratic theory carries an energy $k_B T$ distributed equally in the phase and density fluctuations. Notably, the $1/k^2$ scaling of the wave amplitudes has the important effect that the DFs are determined primarily by the contributions from low momentum fluctuations.

\begin{center}
\begin{figure}
\includegraphics[width = 3.4in]{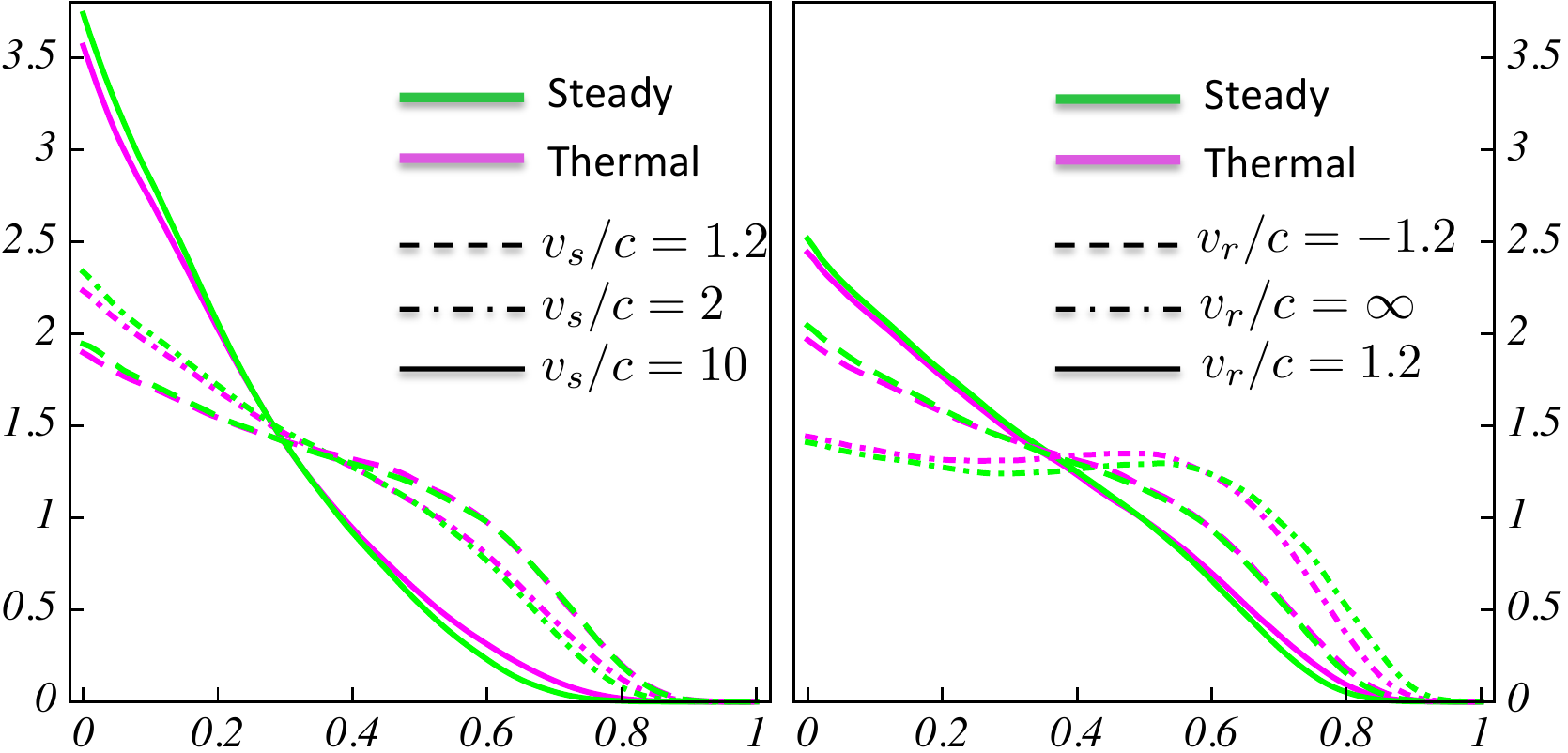}
\caption{Comparison of dynamical contrast distributions in steady/long time limit with thermal distributions at temperatures predicted by Eq. (\ref{eq:Tans}) for (a) equal-time correlations (fixed $v_r/c = \infty$) and different $v_s$, and (b) unequal-time correlations with fixed $v_s/c = 4$. In all plots, system size  $L = 400 \xi_c$, integration length $l = 20 \xi_c$, Luttinger parameter $K = 10$. Dynamical distributions are calculated at $t = 2 l/c$, which is long enough for prethermalization to have occurred.  \vspace{-0.8 cm}}
\label{fig:plots}
\end{figure}
\end{center}

Although $\chi_d$ and $\chi_T$ (in Eqs. (\ref{eq:pdist}) and (\ref{eq:chit})) are ostensibly different, the DF resulting from $\chi_d$ reaches a steady state resembling the thermal DF realized from $\chi_T$ at long times $t \sim l/c$, $l$ being the integration length. To see this, first note that evaluating different moments $\avg{\alpha^m}$ from $P(\alpha)$ corresponds to evaluating correlators of the form $\avg{\prod_i \chi_d (x_i,t_i)}$ and integrating over $x_i$'s. The averages here are taken over the measure $\Pi_k \int dr_k r_k e^{- r^2_k/2} \int d \theta_k / 2 \pi$. Thus, to compare the DF of our dynamical problem to a thermal DF, we compare these correlators of $\chi_d$ to those of $\chi_T$. Calculation (see SI \ignore{\ref{sec:solve}}) reveals that in the long-time limit, these correlators, factorize pairwise into correlators which can be characterized by an amplitude $f_d(k)$ in analogy to the amplitude $f_T(k)$ of Eq. (\ref{eq:ftk}) that arises in the thermal case with
%
%
%To see this, we examine the correlators of $\chi_d(x,t_0)$ in the long time limit ($t_0 \rightarrow \infty$) and compare them with correlators of $\chi_T(x)$. Our aim is to prove the following
%\be
%\lim_{T \rightarrow \infty} \int^{2 T}_{T} \frac{dt_0}{T} \avg{\prod_i \chi_d (x_i,t_0)} = \avg{\prod_i \chi_T (x_i)}
%\ee
%where all averages are taken over the measure $\int dr_k r_k e^{- r^2_k/2} \int d \theta_k / 2 \pi$. The expression on the LHS formalizes the description of the `long time limit' - for long times $t_0$, all integrals over $k$ in $\chi_d (x,t_0)$ have essentially a time averaging effect. 
%Evaluating different moments of $P(\alpha,t_0)$ corresponds to evaluating such correlators of $\chi_d(x_i,t_0)$ and integrating over $x_i$. Thus, to compare the DFs, we compare these correlators. 
%
\be
f_d (k) = \frac{g}{ c \abs{k}} \frac{g \rho}{c \abs{k}} (a_R + a_L) \left( 1 + \gamma^2_r \left( \frac{\xi_c k}{2\pi} \right)^2 \right). 
\label{eq:fdk}
\ee
Here $a_R = \eta_s (1 - c/v_r)$ and $a_L = (1 + c/v_r)/\eta_s$  are the effective dilation factors of right and left moving waves as discussed above and $\gamma^2_r = 1/a_R a_L$. We see that the amplitudes $f_T (k)$ and $f_d (k)$ ( in Eqs. (\ref{eq:ftk}) and (\ref{eq:fdk})) are of the same form, but for an extra $k$-independent factor. This factor gives a UV dependent contribution to correlations, which are small (cf. numerical simulations in Fig. (\ref{fig:plots}) and SI \ignore{\ref{sec:nsim}})- for example, its effect on the expectation value $\avg{e^{i \phi(x_1,t_1) - \phi(x_2,t_2)}}$ is a multiplicative factor of order $e^{-1/4K (1/a_R + 1/a_L)} \sim 1 $, for Luttinger parameter $K \gg 1$. Ignoring this extra term, and comparing the results for the dynamical and thermal amplitudes, we arrive at the most significant result of this Letter - the temperature $T$ characterizing the prethermalization of equal and unequal time correlations is 
\be
T(v_s, v_r) = \gamma_s \left(1 - \frac{c^2}{v_s v_r} \right) T_0 \; \; \; ; \; \; \; T_0 = \frac{g \rho}{2},
\label{eq:Tans}
\ee
where $\gamma_s = 1/\sqrt{1- u^2_s / c^2} = 1/\sqrt{1 - c^2/v^2_s}$ is the Lorentz factor corresponding to the boost velocity $u_s$. We first examine some general features of this result. In the limit $v_s \rightarrow \infty$, $v_r \rightarrow \infty$, that is, measurement of equal-time correlations when the condensate is split suddenly, we expectedly reproduce the sudden quench result $T = T_0 = g\rho/2$. In the limit $v_r \rightarrow \infty$, but finite $v_s$, which corresponds to the measurement of equal time correlations of a body moving with velocity $u_s = c^2 / v_s$, we find that $T = \gamma_s T_0 = T_0 (\eta_s + 1/\eta_s)/2$, which implies that the moving body seems hotter, by a Lorentz factor, in agreement with the conclusions of Ott \cite{Ott}. Interestingly, for the case $v_r = v_s$, we obtain the Planck result \cite{Planck} - $T = T_0 / \gamma_s$, which indicates that measurements isochronous in the rest frame of the moving body, appear colder. In the general case of finite $v_s >c $ and $\abs{v_r} > c$,we find that the temperature $T$ satisfies the relation $T_0/\eta_s \le T \le \eta_s T_0$. In particular, the limits $T = \eta_s T_0$ and $T =T_0/\eta_s$ are reached for the cases $v_r = -c$ and $v_r = c$ respectively - this is a consequence of the fact that at such values of $v_r$, we simply measure the correlations of a single set of waves, right-moving or left-moving that correspond to the temperatures $\eta_s T_0$ and $T_0/\eta_s$ respectively.
%\begin{center}
%\begin{figure}
%\subfloat[]{\includegraphics[width = 1.7in]{equal2.pdf}} \subfloat[]{\includegraphics[width = 1.7in]{unequal2.pdf}}
%\caption{Comparison of dynamical contrast distributions in steady/long time limit with thermal distributions at temperatures predicted by Eq. (\ref{eq:Tans}) for (a) equal-time correlations (fixed $v_r/c = \infty$) and different $v_s$, and (b) unequal-time correlations with fixed $v_s/c = 4$. In all plots, system size  $L = 400 \xi_c$, integration length $l = 20 \xi_c$, Luttinger parameter $K = 10$. Dynamical distributions are calculated at $t = 2 l/c$, which is long enough for prethermalization to have occurred.  \vspace{-0.5 cm}}
%\label{fig:plots}
%\end{figure}
%\end{center}
\paragraph*{Accelerating Probes} A simple extension of the measurement protocol employed so far would be to release the condensates in an accelerated fashion, instead of a constant velocity $v_r$. One could then ask if additional conformal symmetries of the Luttinger liquid allow us to interpret such accelerated unequal time correlations as thermal correlations at some effective temperature. An answer to such a question, was first provided by Unruh \cite{Crispino}, who showed that a relativistically accelerating detector traveling through vacuum (on a trajectory $x^2 = (ct)^2 + c^4/a^2$, with proper acceleration with $a$)  finds a thermal flux of particles at temperature $T_U = a/ (2 \pi c)$. 
Unfortunately, the Unruh trajectory is time-like, and hence correlations along such a trajectory cannot be obtained in the interference experiments we consider.  
%Secondly, the Unruh effect is a $local$ effect \cite{local} (see SI \ref{sec:local}) - the notion of an acceleration temperature applies only to local (in conformal coordinates) field observables. 
To circumvent this difficulty, we suggest a simple modification of the Unruh trajectory to the following : $x^2 = (ct)^2 - c^4/a^2$ (see SI \ignore{\ref{sec:local}}). This trajectory is space-like, and correlations along such a trajectory are measurable in experiments. Moreover, we can immediately see how an analog of the Unruh result shows up in such a case. First, we define the conformal coordinates $\{\xi, \eta_s\}$ according to the transformation relations - $x = \frac{1}{a} e^{a \eta_s} \sinh{ a \xi}$, $t = \frac{1}{a} e^{a \eta_s} \cosh{ a \xi} $. In such coordinates, our modified Unruh trajectory is simply $\eta_s = 0$. Now, when we express the correlations of a $zero$ temperature Luttinger liquid in these conformal coordinates, we see that they look thermal, that is, $\avg{\phi(\xi, \eta_s = 0) \phi(\xi', \eta_s = 0)}_{T = 0} \propto \log \left( \sinh^2 (a \frac{ \xi - \xi'}{2 c^2}) \right) \approx a/c^2 \abs{\xi - \xi'}$, with a temperature $T = T_U = a / 2 \pi c$, (see SI \ignore{\ref{sec:unruh}}) which is the Unruh result. The experimental protocol to measure this effect is therefore - (i) to split the condensate slowly, preparing the phase difference field in a very low temperature \cite{optimalsplit} state with $T < 2 \pi c / L$ and $L$ is the length of the condensate and (ii) measure the correlations along the specified trajectory, expressing the result in the conformal coordinate $\xi$. 

\paragraph*{Discussion} We first discuss the implications of our findings on the debate \cite{Callen} over the Lorentz transformation law of temperature. As mentioned earlier, the effective temperature $T$ (Eq. (\ref{eq:Tans})) characterizing unequal time correlations (characterized by some fixed velocity $v_r$) spans a range $T_0 / \eta_s \le T \le \eta_s T_0$. In the case of a body at rest, $\eta_s = 1$ ($u_s = 0$, $v_s = \infty$), and this range of temperatures collapses to a single result $T_0$. This makes the notion of temperature of a stationary body meaningful -- all correlations (equal or unequal time) correspond to the same fixed temperature. For a moving body, the mere existence of correlations exhibiting different effective temperatures indicates that providing a $definitive$ answer to the Lorentz transformation law of temperature is not possible \cite{Callen}. 

This range of measured temperatures is analogous to the observation of dipole anisotropy in the temperatures of the CMB. The CMB itself has a preferred frame, in which the radiation is postulated to be isotropic and thermal. Due to the motion of our galaxy, telescopes pointing in different directions sense this radiation to have a range of Doppler-shifted temperatures \cite{DipoleCMB} in a manner analogous to the right and left moving waves considered here. In our problem, this range of detectors is encoded in the choice of different sets of unequal time correlations characterized by a release velocity $v_r$. A nice feature of an ultra-cold atoms experiment over the measurement of the CMB, of course, is the possibility of tuning $v_s$ and $v_r$, which gives a range of anisotropies and temperatures. 

Finally, we remark that our methods can be generalized to analyze problems of quenches along more complicated  supersonic trajectories. The Luttinger model that governs the post-quench dynamics possesses conformal symmetries in addition to the Lorentz symmetry we have considered here. Thus, cases of more complicated quench trajectories can be considered analogously by mapping the problem to a sudden quench in appropriately chosen conformal coordinates.

\paragraph*{Acknowledgements} We thank Koenraad Schalm for useful discussions. K.A., E.D.T. and E.D. acknowledge support from Harvard-MIT CUA, DARPA OLE program, the ARO-MURI on Atomtronics, ARO MURI Quism program. B.R., T.L. and J.S. acknowledge the support by the EU through the projects SIQS (GA\#284584) and QuantumRelax (ERCADG-320975). B.R. thanks the Austrian Science Fund (FWF) Doctoral Program CoQuS (W1210).

\bibliographystyle{apsrev4-1}
\bibliography{cond1}

\vspace{1cm}

\begin{center} \textbf{SUPPLEMENTAL INFORMATION} \end{center}

\section{Experimental Proposal} 
\label{sec:protocolexp}

The proposed scheme could be realized using a matter-wave interferometer on an atom chip. In the following, we outline the technical details of this implementation.

Atom chips \cite{atomchipprl,Reichel2011} offer a versatile platform for the manipulation of ultracold 1D Bose gases via near-field radio-frequency (RF) dressing. This enables the implementation of a large variety of adiabatic dressed-state potentials \cite{RFNature,RFgen} for the atoms. For example, by appropriately choosing the properties of the RF radiation, an initially harmonic confinement can be dynamically deformed into a double well potential, thereby realizing an experimentally robust transverse splitting of a single 1D Bose into two parts~\cite{Schumm2005}. 

In this splitting process, the amplitude of the RF radiation can be used to tune the height of the barrier that separates the two potential wells, thus enabling full control over the tunnel coupling between the two parts of the system. Consequently, engineering a gradient of the RF amplitude over the length of the 1D cloud can be used to realize a position dependent splitting process. 

\begin{figure*}
  \centering
  \includegraphics[width=0.7\textwidth]{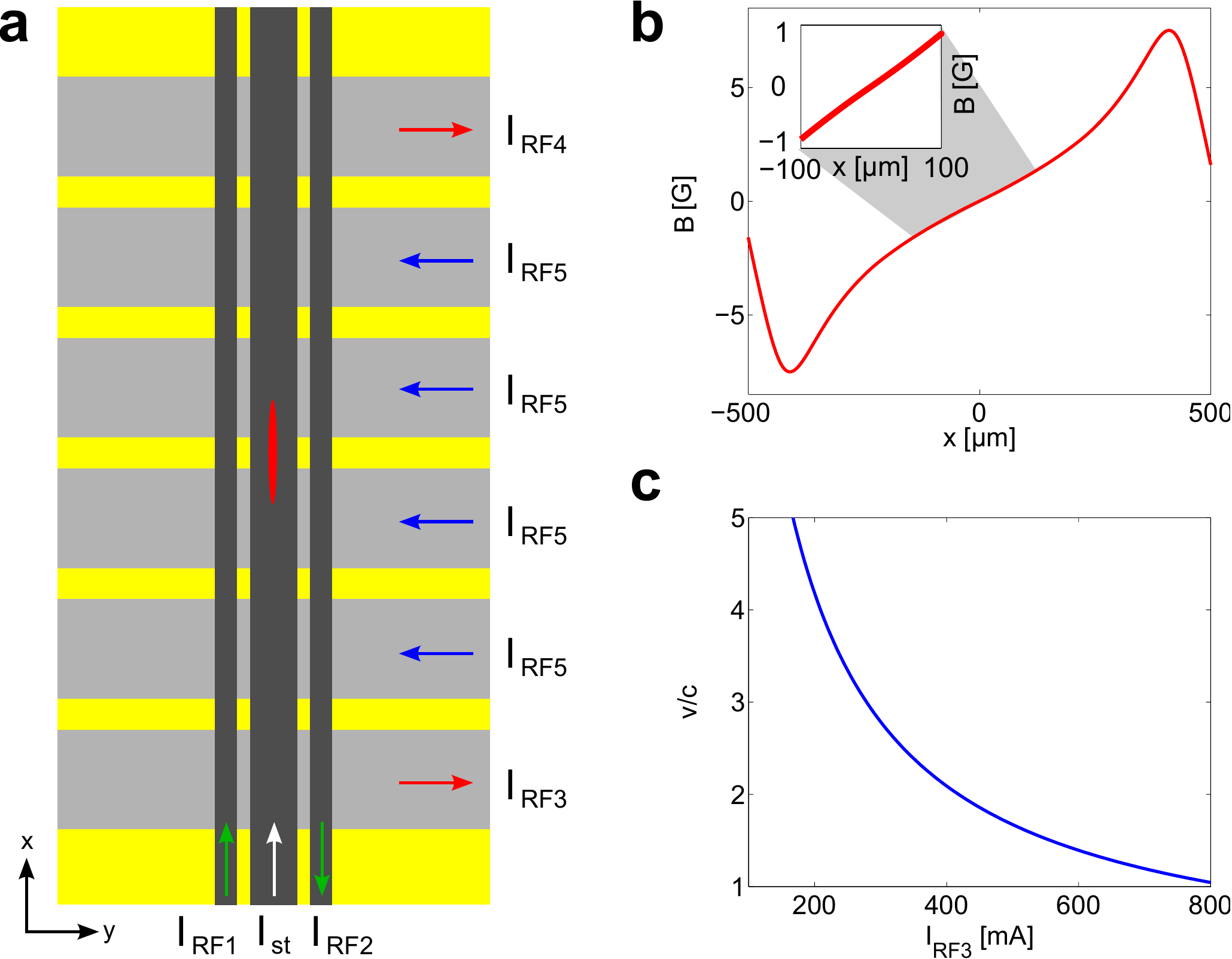}
  \caption{{\bf (a)} Schematic layout of the atom chip. A double-layer chip is used to trap and manipulate a single 1D Bose gas (red, not to scale) by applying various static (white arrow) and radio-frequency (RF) currents (red, green and blue arrows)~\cite{Trinker2008}. The static trap is formed using the current $I_{st}$ in the main trapping wire and an external bias field; additional currents in perpendicular wires can be used to provide longitudinal confinement (not shown for clarity). The primary RF currents $I_{RF1}$ and $I_{RF2}$ (green arrows) are applied through wires adjacent to the main trapping wire. A spatially varying secondary RF field is created using the currents $I_{RF3}$ and $I_{RF4}$ (red arrows). The much smaller currents $I_{RF5}$ (blue arrows) in the four (or more) central wires compensate a spurious field component in the longitudinal direction. 
The secondary RF is superposed in-phase with the spatially homogeneous primary RF field to realize an RF gradient at the position of the atomic cloud. 
{\bf (b)} Example RF gradient realized using 200\,$\mu$m wide perpendicular wires carrying $I_{RF3}$ = $I_{RF4}$ = 650\;mA and
$I_{RF5}$ = 10\;mA. The inset demonstrates the linearity over the typical extension of the cloud, with $dB_2/dz$ = 1\,G/100\,$\mu$m. The RF gradient also leads to a potential tilt. For $^{87}$Rb in the F\,=\,2, m$_F$\,=\,2 state, this tilt is on the order of 200\,kHz/100\,$\mu$m, which can be compensated by a linear electric field gradient $dE/dz$=5\,mV/$\mu$m$^2$, with moderate maximum fields on the order of 0.3\,V/µm. 
{\bf (c)}~Example for a trap with $\omega_{\parallel} = 2\pi \cdot 15$\,Hz, $\omega_{\perp} = 2\pi \cdot 2$\,kHz, containing 6000 atoms. The splitting velocity $v$ is tunable via the applied secondary RF current and reaches values close to the speed of sound.
}
  \label{fig:main_fig}
\end{figure*}

In Fig.~\ref{fig:main_fig} we present an example of a chip configuration to implement such an RF gradient. In the suggested configuration, the atoms are trapped in a standard Ioffe-Pritchard-type microtrap that is created by combining the static field of a straight trapping wire with an external magnetic bias field~\cite{Reichel2011}. Longitudinal confinement is provided by using additional wires that are oriented perpendicular to the main trapping wire. RF radiation is applied via a pair of wires that are located adjacent to the central static trapping wire. For a $\pi$ phase shift between the currents in these two wires the resulting RF field is linearly polarized in the vertical direction, leading to a horizontal double well potential. Increasing the amplitude of this RF radiation realizes a rapid and homogeneous splitting of the atomic cloud, as demonstrated for example in Ref.~\cite{Gring2012}.

To implement the spatial gradient of the RF amplitude a secondary RF field can be applied via two additional wires, which are oriented perpendicular to the main trapping wire. Co-propagating RF currents in these wires create an RF field where the main component is collinear with the field vector of the primary RF field. The vertical amplitude of this secondary field changes approximately linearly along the length of the cloud, as shown in 
Fig.~\ref{fig:main_fig}. Any unwanted component of this field in the longitudinal direction, which would lead to a position dependent tilt of the double well, can be strongly suppressed using additional perpendicular wires (see Fig.~\ref{fig:main_fig}). Furthermore, the linear RF gradient leads to a tilt of the trapping potential, which could be compensated by a linear electric field gradient. Such a field gradient can be created using charged elements on the chip~\cite{Kruger2003}. 

Superposing the secondary RF field in-phase with the primary RF field creates different RF amplitudes along the cloud. Ramping up the amplitude of the primary RF therefore results in a decoupling point that moves through the system at a finite velocity $v_s$.

The value of this velocity is determined by the ratio of the temporal gradient $dB_1/dt$ of the primary RF field and the spatial gradient $dB_2/dx$ of the secondary RF field, where $B_1$ and $B_2$ denote the respective RF field amplitudes. Note that locally the splitting is still determined by $dB_1/dt$, which can easily be made fast enough to realize a quasi-instantaneous local splitting. The gradient $dB_2/dx$ can then be chosen accordingly to achieve splitting velocities close to the speed of sound.  The effective temperature of the prethermalized state can then be measured through the phase-correlation function obtained via matter-wave interferometry~\cite{Schumm2005,Gring2012,Langen2013}. 

In such measurements, unequal time correlation functions can be accessed by outcoupling small fractions of the gas from the trap using Raman or RF transitions along a chosen trajectory. The individual interference patterns of the released atoms can then be measured in time-of-flight expansion using single-atom-sensitive light-sheet imaging \cite{Buecker}. This ensures a minimal perturbance to the system during its ongoing evolution, as (a) only a small fraction of the atoms has to be outcoupled and (b) the imaging process does not influence the in situ cloud. A particular `release' velocity, $v_r$,  can then be chosen by correlating the phases from temporally separated interference pictures.

%%%%%%%%%%%%%%%%%%%%%%%%%%%%%%%%%%%%%%%%%%%%%%%%%%%%%%%%%%%%%%%%%%%%%%%%%%%%%%%%
%%%%	Figure: main figure
%%%%%%%%%%%%%%%%%%%%%%%%%%%%%%%%%%%%%%%%%%%%%%%%%%%%%%%%%%%%%%%%%%%%%%%%%%%%%%%%

\section{Initial Correlations}
\label{sec:diag}

The ground state of the Hamiltonian $H(-\infty)$ describes the initial correlations of the system. In Fourier space 
\be
H(-\infty) = \int dk \; \left( \frac{\rho}{4 m} k^2 + J \right) \phi_k \phi_{-k} + g n_{k} n_{-k} ,
\ee
and can be diagonalized by introducing operators bosonic operators $b_k = \alpha_k \phi_k + i \beta_k n_k$, with $\alpha^2_k = \omega_k / 4 g$, $\beta^2_k = g / \omega_k$ related to the energy eigenvalues $\omega_k = \sqrt{4g \left( J + \frac{\rho k^2}{4 m} \right) } $. In the case $J = 0$, we see that the dispersion becomes linear in $k$ with a velocity $c = \sqrt{ \rho g/m}$. This is the sound velocity of the Luttinger liquid, and the speed which determines the light-cone within which response functions can be non-zero. We can now evaluate the correlations of the number field $n$
\be
\avg{n(x)n(x')} = \int \frac{ dk }{ 2 \pi} \frac{1}{4 \beta^2_k} e^{i k (x-x')}.
\ee
For $ k^2 \ll 4 m J / \rho$, we can neglect the dispersion of $\omega_k$, and find $\avg{n(x)n(x')} \approx \sqrt{J/ 4g} \delta(x-x')$, where the $\delta$ function is naturally smeared over the length scale $2 \pi \sqrt{ \rho/ 4 m J} = 1 / \xi_c$. Thus, to get the desired result for the initial correlations, we simply set $J = g \rho^2$. 

\section{Intricacies associated with Lorentz Boosting}
\label{sec:fin}
\begin{center}
\begin{figure}
\includegraphics[width = 3in]{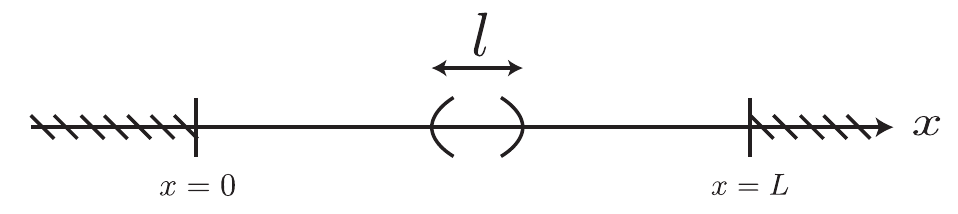}
\caption{Fictitious perturbations coming from outside the condensate ($x<0$, $x>L$) do not affect measurements in the region $x = x_0$ to $x = x_0 + l$ as long as the experimental run time $t < x_0/c$ and $t < L/v_s + (L - x_0 -l )/c$. For small region in the center of a condensate, $x_0 \sim L/2$, the time of validity of the theory is long enough for prethermalization (which occurs on a timescale $\sim l/c$ after the quench passes the region of interest) to take place. } 
\label{fig:fin}
\end{figure}
\end{center}

Given that experiments really work with finite size systems, the more accurate representation of the problem would be in terms of the following action
\be
S_{fin} = \int^{L}_{0} dx \int^{\infty}_{-\infty} dt\frac{1}{4 g} (\partial_t \phi)^2 - \frac{\rho}{4m} (\partial_x \phi)^2 - J \Theta_s( x- vt) \phi^2 .
\label{eq:b1}
\ee
The problem with this action however, is that the finite length of the system clearly breaks Lorentz symmetry (even without involving the splitting perturbation). Furthermore, the Lorentz boost creates an action in the new space-time variables $\{ x',t' \}$ wherein the range of $x'$ is time $t'$ dependent. This is clearly a unwelcome complication. To alleviate this problem, we choose to work with the extended action, where $x$ ranges from $-\infty$ to $\infty$.  
This action is not simply a spatially extended version of the action in Eq. (\ref{eq:b1}). In this action, perturbations start from time $t = -\infty$, and carry on long after the perturbation is supposed to have left the system. Thus, in principle, this action allows for `fictitious' perturbations from outside the true condensate boundaries to interfere with the results that would be produced by analyzing the action in Eq. (\ref{eq:b1}). But owing to the light cone physics of response functions of the superfluid action, one can indeed compute correlation functions for the finite system from this extended action, $provided$ we limit ourselves to evaluating field operators at points in space-time that are not affected by these fictitious perturbations (see Fig. \ref{fig:fin}). Immediately, this implies that we can only work with this extended action approach only when the perturbation velocity $v$ is greater than the superfluid sound velocity $c$. For the opposite case of $v_s < c$, the slicing sets up ripples that propagate at speeds faster than the slicing perturbation. Consequently, many perturbations from beyond the length of the condensate, which we deem fictitious, reach it even before the slicing perturbation does. 

\section{Solving for Distribution Functions}
\label{sec:solve}

The dynamics of the system in the boosted frame is given in terns of the Hamiltonians $H'_{t'<0}$ and $H'_{t'>0}$ given in Eq. (\ref{eq:Hb}) of the main text. To solve for the dynamics, we diagonalize these Hamiltonians as follows - $H'_{t'<0} = \int dk \; \omega_{k,1} b^\dagger_{k,1} b_{k,1}$ and $H'_{t'>0} \int dk \; \omega_{k,2} b^\dagger_{k,2} b_{k,2}$, where for our purposes, it is crucial to know that the relation between these sets of bosonic operators is given as $b_{k,2} = A^+_k b_{k,1} - A^-_k b^\dagger_{-k,1}$ with $A^\pm_k = \frac{1}{2} \left( \sqrt{\frac{2 g \rho}{c \abs{k}}} \pm \sqrt{\frac{c\abs{k}}{2 g \rho}}\right)$. The time evolution of the field $\phi'(x',t')$ for $t'>0$ is given in terms of bosons $b_{k,2}$ as
\be
\phi'(x',t') = \int \frac{dk}{\sqrt{2 \pi}} \sqrt{\frac{ g}{c \abs{k}}} \left( b_{k,2} e^{i (kx' - \omega_{k,2} t')} + h.c. \right).
\ee

To find correlation functions of these field operators, it is useful to represent them in terms of $b_{k,1}$ bosons as the wave function is a vacuum of these bosons. The evolution in terms of these operators is 
\begin{align}
\phi'(x',t') &= \int{dk} \left( \gamma_k(x',t') b_{k,1} + \gamma^*_k(x',t') b^\dagger_{k,1} \right), \nonumber \\
\gamma_k (x',t') &= \sqrt{\frac{ g}{2 \pi c \abs{k}}} \left( A_k e^{-i c \abs{k} t'} - B_k e^{ i c \abs{k} t'} \right) e^{i k x'}.
\label{eq:gamma_sk}
\end{align}

We want to evaluate distribution functions of the integrate phase contrast $\alpha = \abs{\int dx/l e^{i \phi(x,t)}}^2$. To do so, we need correlations of the form - $\avg{\prod_i e^{\epsilon_i i \phi'(x'_i,t'_i)}}$, with $\epsilon_i = \pm 1$, and the points $\{ x'_i, t'_i \}$ are space-like separated. Now, because these points are space-like separated, the field operators at these points commute, and the product over the exponentials of these operators can be replaced by a sum in the exponential of these operators. In terms of the $\gamma_k$'s described in Eq. (\ref{eq:gamma_sk}), we find 
\be
\avg{e^{i \sum_i \epsilon_i \phi'(x'_i, t'_i)}} = \prod_k e^{-\frac{1}{2} \abs{ \sum_i \epsilon_i \gamma_k (x'_i,t'_i)}^2} .
\label{eq:corr1}
\ee
Now, to evaluate the contrast to some power $m$, $\alpha^m$, we will have $i$ ranging from $1$ to $2 m$, set half of the $\epsilon_i = 1$ and the rest to $-1$, and integrate over each of these variables $x'_i$ (More precisely, we will be performing an integral over $x_i$, that is, over a lab frame coordinate directly related to $x'_i$). It is not immediately apparent that each of these integrals over $x'_i$ can be performed independently of one another. But a Hubbard Stratanovich trick helps us separate the integrals. A single $k$ term in the above result in Eq. (\ref{eq:corr1}) can be expressed as
\begin{align}
& e^{-\frac{1}{2} \xi_k \xi^*_k} = e^{- \frac{1}{2} \left( (\Re{\xi_k})^2 + (\Im{\xi_k})^2 \right)} \nonumber \\
&= \int^{\infty}_{-\infty} \frac{d \lambda_{k,1}}{\sqrt{2\pi}} \int^{\infty}_{-\infty} \frac{d \lambda_{k,2}}{\sqrt{2\pi}} e^{- ( \lambda^2_{k,1} + \lambda^2_{k,2} )/2 } e^{ i \lambda_{k,1} \Re{\xi_k} + i \lambda_{k,2} \Im{\xi_k}} ,
\end{align}  
with the expressions for $\Re \xi_k$ and $\Im \xi_k$ 
\begin{align}
\Re{\xi_k} &= \sqrt{\frac{g}{2 \pi c \abs{k}}} \sum_i \sum_{a = \pm} a \epsilon_i \left( A^a_k \cos{(k x'_i - a c \abs{k} t'_i)} \right), \nonumber \\
\Im{\xi_k} &= \sqrt{\frac{g}{2 \pi c \abs{k}}} \sum_i \sum_{a = \pm} a \epsilon_i \left( A^a_k \sin{(k x'_i - a c \abs{k} t'_i)} \right). \nonumber \\
\end{align}
Now the integrals over $x'_i$ can be performed independently. Before expressing the final results, we make some modifications - we again perform a change of variables, defining $r^2_k = \lambda^2_{k,1} + \lambda^2_{k,2} $ and $\lambda_{k,2} = r_k \cos {\theta_k}$. And finally, we transform our results from the $\{x',t'\}$ to the laboratory frame coordinates $\{x,t \}$, which results in the introduction of the doppler factor $\eta_s$ through the transformation laws - $x' - ct' = \eta_s (x - ct)$ and $x' + ct' = (x+ct)/\eta_s$. Also, the time variable $t$ in these expressions can be set to any function of $x$ as long as the points $\{ x_i, t_i\}$ remain space-like separated, to get results for unequal time correlations. 
In summary, we find 
\be 
\avg{\abs{\int \frac{dx}{l} e^{i \phi(x,t)}}^{2m}} = \prod_k \int \frac{d \theta_k}{2\pi} r_k d r_k e^{- r^2_k / 2} \abs{\int \frac{dx}{l} e^{i \chi_d}}^{2m} ,
\ee
where the result for $\chi_d$ is as expressed in the Eq. (\ref{eq:pdist}) of the main text. Moreover, using the fact that $\int d\alpha \alpha^m P(\alpha) = \avg{\abs{\int \frac{dx}{l} e^{i \phi(x,t)}}^{2m}}$, we arrive at the result for probability distribution function $P$ as also given in Eq. (\ref{eq:pdist}) of the main text. 

To compare these distribution functions of the dynamical problem, we also need to formulate the result for the distribution functions of the contrast $\alpha$ for independent thermal condensates. One can follow an entirely similar approach as that adopted above, and find that, for the thermal correlations, we find that $\chi_d$ is replaced by a thermal version $\chi_T$ given by the relation 
\be
\chi_T = \int \frac{dk}{\sqrt{2 \pi}} r_k \sqrt{\frac{g }{c \abs{k}} \coth{\frac{c \abs{k}}{2 k_B T}}} \sin \left(k x - c \abs{k} t + \theta_k \right) ,
\label{eq:chithermal}
\ee
wherein we set $t = 0$ (the choice of $t$ is unimportant in the thermal case) to get the result in Eq. (\ref{eq:chit}) of the main text. 

As mentioned in the main text, we compare the DFs obtained in the dynamical problem and the thermal case, by compare the correlators of the functions $\chi_T$ and $\chi_d$ in the long time limit. Mathematically, our aim is to show the following - 
%To see this, we examine the correlators of $\chi_d(x,t_0)$ in the long time limit ($t_0 \rightarrow \infty$) and compare them with correlators of $\chi_T(x)$. Our aim is to prove the following
\be
\lim_{T \rightarrow \infty} \int^{2 T}_{T} \frac{dt_0}{T} \avg{\prod_i \chi_d (x_i,t_i = t_0 + x_i/v_r)} = \avg{\prod_i \chi_T (x_i)},
\label{eq:corrcomp}
\ee
where all averages are taken over the measure $\int dr_k r_k e^{- r^2_k/2} \int d \theta_k / 2 \pi$. The expression on the left hand side of Eq. (\ref{eq:corrcomp}) formalizes the description of the long time limit - for long times $t_0$, all integrals over $k$ in $\chi_d (x,t_0)$ have essentially a time averaging effect. Evaluating different moments $\avg{\alpha^m}$ from $P(\alpha,t_0)$ corresponds to evaluating such correlators of $\chi_d(x_i,t_0)$ and integrating over $x_i$. Thus, to compare the DFs, we compare these correlators. 

Using the basic result that $\lim_{t \rightarrow \infty} \sin(\theta_1 + k_1 c t) \sin(\theta_2 + k_2 c t) = - \frac{1}{2} \delta_{k_1,-k_2} \cos{(\theta_1 + \theta_2)} + \frac{1}{2} \delta_{k_1,k_2} \cos{(\theta_1 - \theta_2)}$, and perform integrals over the measure $\Pi_k \int dr_k r_k e^{- r^2_k/2} \int \frac{d\theta_k}{2 \pi}$, it is easy to show that

\begin{align}
\avg{\chi_T(x) \chi_T(x')} &= \int \frac{dk}{2 \pi} f_T(k) \cos{k (x-x')}, \nonumber \\
\avg{\chi_d(x) \chi_d(x')} &= \int \frac{dk}{2 \pi} f_d(k) \cos{k (x-x')}. 
\label{eq:corr2}
\end{align}

Moreover, due to the integrals over different $\theta_k$, higher correlations such as $\avg{\chi_{d(T)} (x_1) ... \chi_{d(T)} (x_{2n})}$ can be expressed as a sum of products of pairwise correlations $\avg{\chi_{d(T)} (x_i) \chi_{d(T)} (x_j)}$ with different combinations of $x_i$ and $x_j$. Thus, the amplitudes $f_T(k)$ and $f_d(k)$ in correlators of Eq. (\ref{eq:corr2}) contain all information about the correlations in the thermal system, and the dynamical system in the long time limit. The equivalence of these correlators formally proves the equivalence of the distribution functions. 

\section{Locality of the Unruh effect}
\label{sec:local}
The usual description of the Unruh effect \cite{Birrell} falls along the following lines - consider a particle detector that moves with uniform acceleration $a$ (trajectory $x^2 = t^2 + 1/a^2$) through the vacuum of a scalar field residing in Minkowski space. When one calculates the rates of transition between the internal levels of the detector, one finds a detailed balance, that suggests that the detector's internal levels become populated thermally with a temperature $T = a/2 \pi c$. This suggests, in turn, that the detector finds itself in equilibrium with some sort of a thermal bath. 

To explain this seemingly anomalous result, Unruh showed that the definitions of vacuum according to an internal and a non-intertial observer (like the one on a Unruh trajectory) do not agree - in the Unruh case, the vacuum of an inertial observer, happens to correspond to a thermal state of the Unruh observer. Unruh suggests that an observer proceeding on the trajectory $x^2 = t^2 + 1/a^2$, will see the universe in terms of new conformal coordinates $\{ \xi, \eta_s \}$ related to the Minkowski coordinates $\{ x, t \}$ by the following transformation laws - 
\begin{align}
x &= \frac{1}{a} e^{a \xi} \cosh{ a \eta_s}, \nonumber \\
t &= \frac{1}{a} e^{a \xi} \sinh{ a \eta_s}, \nonumber \\
dx^2 - dt^2 &= e^{2 a \xi} ( d \xi^2 - d \eta_s^2 ) = - d \tau^2 .
\end{align}
It is easy to see that the Unruh trajectory corresponds to a static point in the conformal coordinates, specifically $\xi = 0$. Importantly, at $\xi = 0$, the proper time $\tau$ of the Unruh observer, agrees with the conformal time coordinate $\eta_s$ - this gives this particular conformal transformation special privilege in the eyes of the Unruh detector. 
For the case of a massless scalar field, Unruh argues that the vacuum of the Unruh observer should be defined in terms of positive (in conformal time) frequency modes $u_k = e^{i k \xi - i c \abs{k} \eta_s}$ that satisfy the equation of motion in the conformal coordinates. Unruh then proceeds to show that this vacuum does not agree with the Minkowski vacuum defined in terms of the positive frequency modes $v_k = e^{i k x - i c \abs{k} t} $.  

The essential aspect of the result is the following - it is only the observer at $\xi = 0$ who thinks that the system is thermally populated at the specified temperature $T = a/2\pi c$. Other observers, who happen to be on a different static $\xi$ trajectory, will observe a different temperature. Therefore, measuring space-like correlations, accessible to us in experiments, cannot reveal the Unruh effect. Thus, the Unruh effect is a $local$ effect. A simple way to see this, is to look at equal $\eta_s$ correlations of the ground state of a massless scalar field $\phi$. In Minkowski coordinates, these correlations look like $\avg{\phi(x,t) \phi(x',t')} \sim \ln{ \left( ( x-x')^2 - (t-t')^2 \right)}$. In conformal coordinates, these correlations look like $\avg{\phi(\xi,\eta_s) \phi(\xi',\eta_s)} \sim \ln \left( e^{a \xi} - e^{a \xi'} \right) $. Thus, the correlations are not even translationally invariant in this new frame. 

As described in the main text, to overcome this difficulty, we consider the following new set of conformal coordinates $\{\xi',\eta_s'\}$ obeying transformation laws heavily inspired by Unruh's choice - 
\begin{align}
x &= \frac{1}{a} e^{a \eta_s} \sinh{a \xi}, \nonumber \\
t &= \frac{1}{a} e^{a \eta_s} \cosh{a \xi}, \nonumber \\
dx^2 - dt^2 &= e^{2 a \eta_s} ( d \xi^2 - d \eta_s^2 ) .
\label{eq:transrule}
\end{align}
Here we consider measuring correlations of field operators at the points on the trajectory $\eta_s = 0 : x^2 = (ct)^2 - c^4/a^2$, which is, unlike the Unruh trajectory, a superluminal trajectory. Moreover, $0$ temperature correlations of the field operators in this new coordinates follow $\avg{\phi(\xi,0)\phi(\xi',0)} \propto \ln ( \sinh^2(a \frac{\xi - \xi'}{2} )) \approx a \abs{\xi - \xi'}$ which is the result expected for equal time correlations of the phase field at a large temperature $T \propto a$. 

\section{Distribution functions for accelerating probe experiment}
\label{sec:unruh}

In the case of the experiment designed to measure the Unruh effect, we start out with condensates that are adiabatically separated in a spatially uniform way, so the correlations can be assumed to be thermal at some low temperature $T$. As a first step, we express the result in Eq. (\ref{eq:chithermal}) in terms of the new conformal coordinates $\{ \xi, \eta_s \}$ related to the lab frame coordinates $\{ x,t \}$ using the relations in Eq. (\ref{eq:transrule}). We are interested in correlations on the trajectory $(ct)^2 = x^2 + c^4/a^2$, or equivalently $\eta_s = 0$, and see if these correlations look thermal. As an example, we evaluate the equal (conformal) time correlation $\avg{e^{i (\phi(\xi,\eta_s = 0) - \phi(\xi',\eta_s = 0))}}$, which can be expressed in terms of these $\chi$ functions using the following relation - 
\begin{align}
 &\avg{e^{i (\phi(\xi, \eta_s = 0) - \phi(\xi',\eta_s = 0))}} = \nonumber \\
 &e^{  \avg{\chi_T (\xi, \eta_s = 0) \chi_T (\xi', \eta_s = 0)} - \avg{\chi_T (\xi, \eta_s = 0)}^2},
\end{align}

where the average over the $\chi$ field is over the measure $\frac{d \theta_k}{2 \pi} dr_k r_k e^{- r^2_k/2}$. This result follows directly from methods used in the evaluation of the full distribution functions, and the essence of the result is unchanged - we exchange the field $\phi$ for $\chi$ and a different measure over which we average the result. For $T = 0$, the result can be evaluated exactly with a UV cut-off $k_c = 2 \pi / \xi_c$. We find
\begin{align}
\avg{e^{i (\phi(\xi) - \phi(\xi'))}} &= e^{- \frac{1}{2 K} \ln {\left( \frac{c k_c}{T_U} \right)} } e^{- \frac{2 g T_U}{ c^2 a} \ln{(4 \sinh^2 (a/c^2 (\xi - \xi')/2))}} \nonumber \\
&\approx const \times e^{- \frac{2 g T_U}{c^2} \abs{ \xi - \xi'} }  ,
\label{eq:Unruhtherm}
\end{align}
where we have expressed the result in terms of the Luttinger parameter $K = \rho \xi_c / 2$ and have suggestively introduced $T_U = a / 2 \pi c$. For a thermal system at temperature $T$, It is easy to show that the correlation for large temperatures of the order of the chemical potential,  $\avg{e^{i (\phi(x,t) - \phi(x',t))}} \approx e^{ - \frac{2 g T}{c^2} \abs{x - x'}}$, which is of the same form as the correlation in Eq. (\ref{eq:Unruhtherm}). Therefore, the correlations of the condensates measured along this superluminal trajectory look thermal with a temperature $T  = T_U$ in these conformal coordinates. It is also important to note that besides a thermal looking exponential decay of correlations with (conformal) distance, we also have an additional constant multiplicative factor of $e^{-ln( c k_c / T_U)/2K}$. In order for this factor to not impact the results, we require  $\abs{\ln{ ck_c/T_U}} \ll 2K$, which can be satisfied by a wide range of values for the acceleration. Finally, we remark that the $T = 0$ result is valid in a finite system as long as $T \ll 2 \pi c / L$, such that even the energy modes are not thermally populated, and this is understood to be experimentally feasible. 

\section{Numerical Simulations}
\label{sec:nsim}

To numerically compute the distribution function $ P (\alpha, t)$ using Eq. (\ref{eq:pdist}) of the main text, we perform Monte-Carlo integration over the variables $r_k$ and $\theta_k$ for a finite number of $k = -2 \pi / \xi_c, ...., 2 \pi / \xi_c$ in steps of $2 \pi / L$. All continuous integrals over $k$ undergo the replacement - $\int dk / \sqrt{2 \pi} \rightarrow (1/\sqrt{L}) \sum_k$ to yield the finite size results for our problem. 

\begin{center}
\begin{figure*}
\subfloat[]{\includegraphics[width = 1.7in]{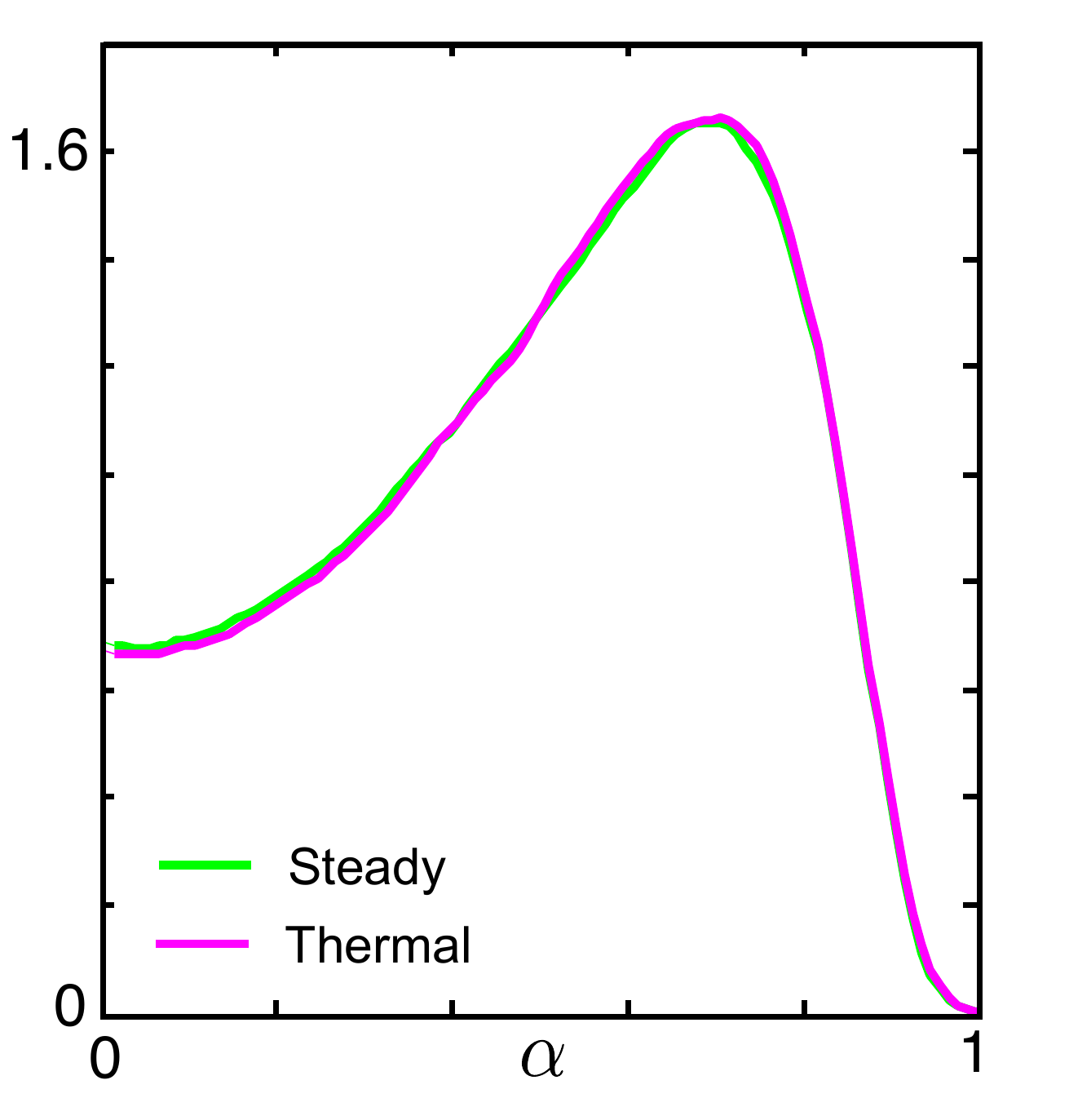}} \subfloat[]{\includegraphics[width = 1.7in]{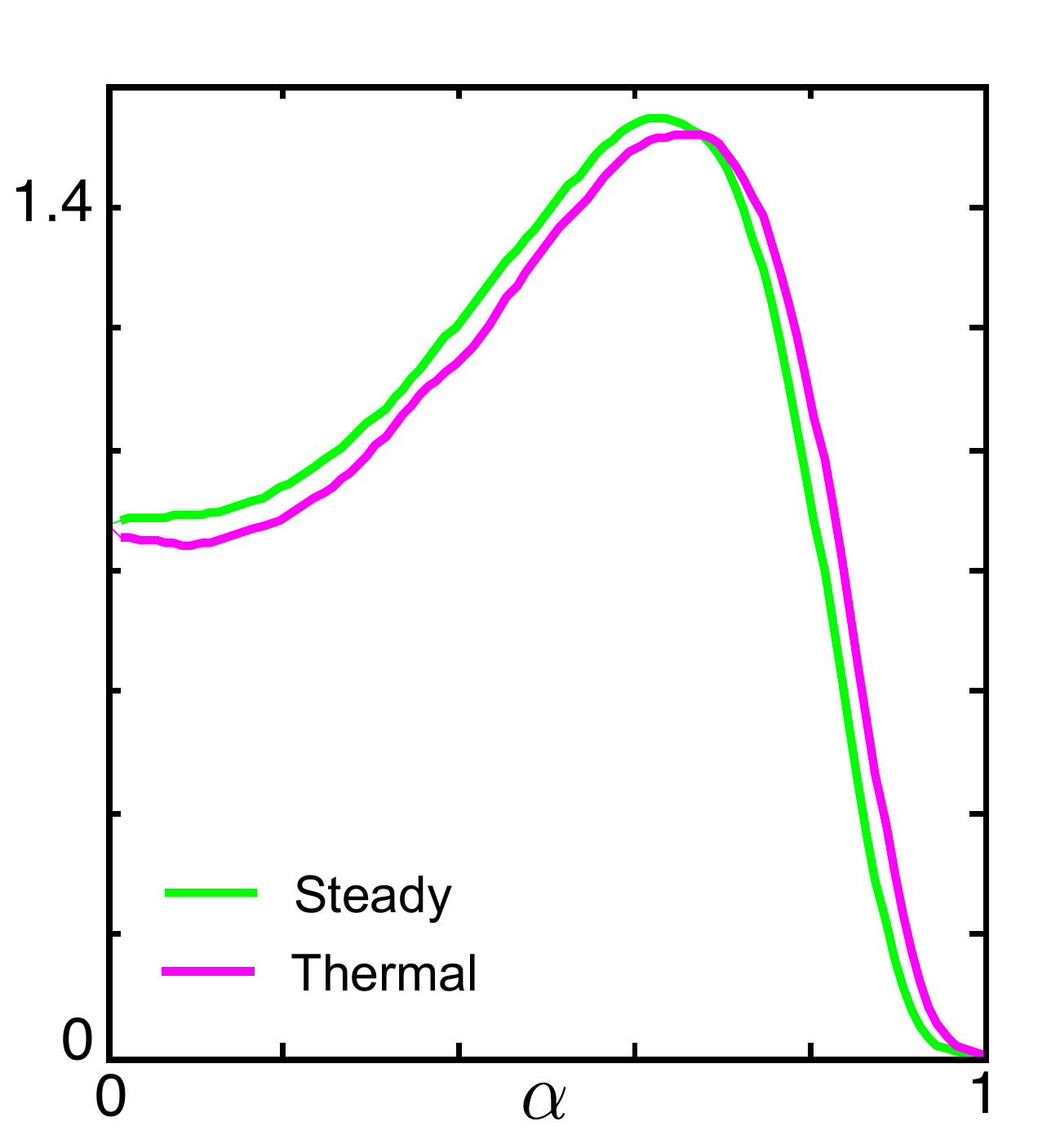}} \subfloat[]{\includegraphics[width = 1.74in]{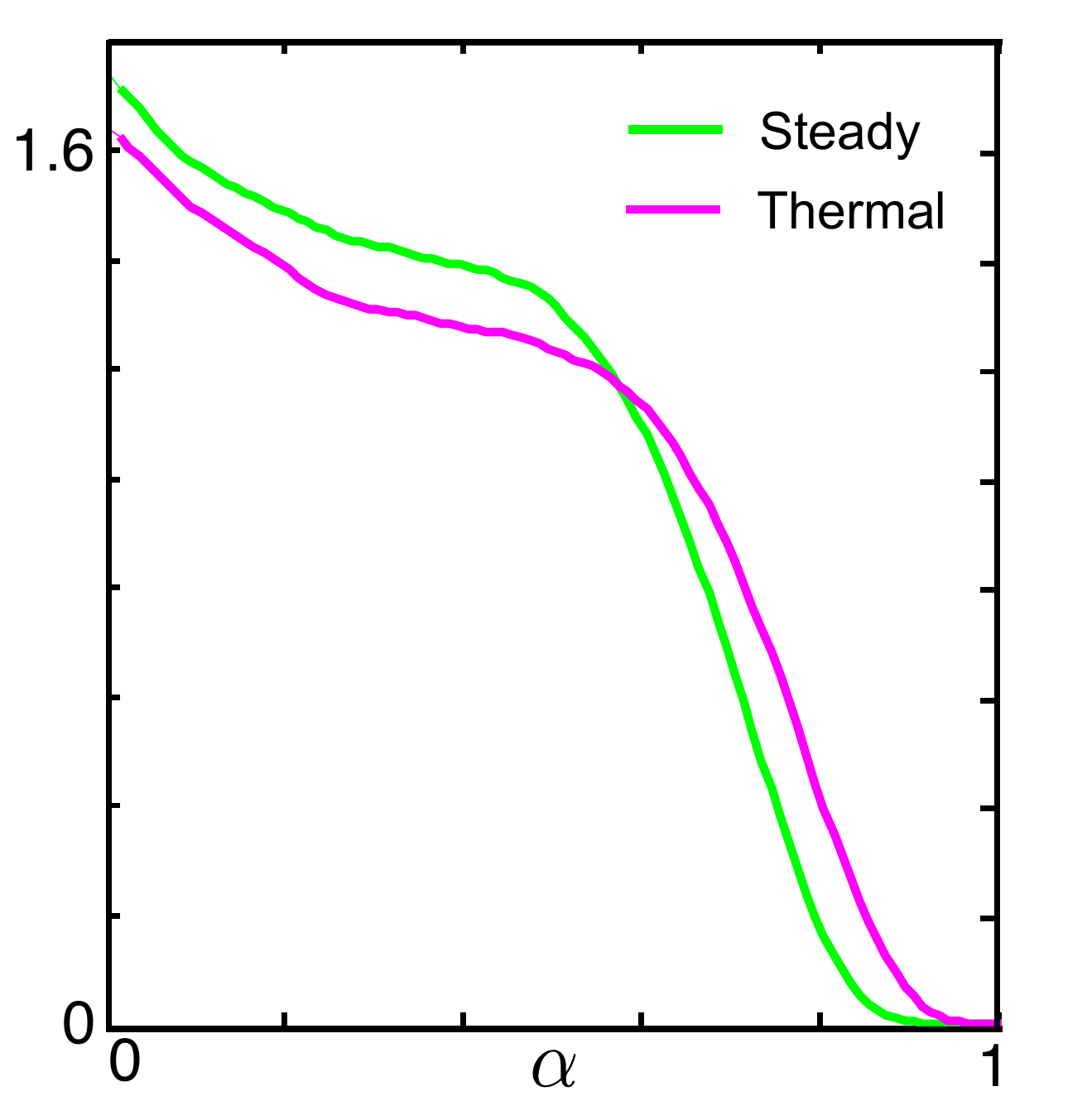}}
\caption{Comparison of dynamical contrast distributions in steady/long time limit with thermal distributions at temperatures predicted by Eq. (\ref{eq:Tans}) of the main text, for splitting velocities (a) $v_s/c = 10$, (b) $v_s/c = 2$ and (b) $v_s/c = 1.2$. In all plots, the system size is $L = 400 \xi$, integration length $l = 10 \xi$ and Luttinger parameter $K = 10$, and the dynamical distributions are measured at $t = 40 \xi_c / c$, which is long enough for prethermalization to have occurred.  For slow velocities, deviations between the dynamical and thermal distribution functions increases.}
\label{fig:equal1}
\end{figure*}
\end{center}

\begin{center}
\begin{figure*}
\subfloat[]{\includegraphics[width = 1.74in]{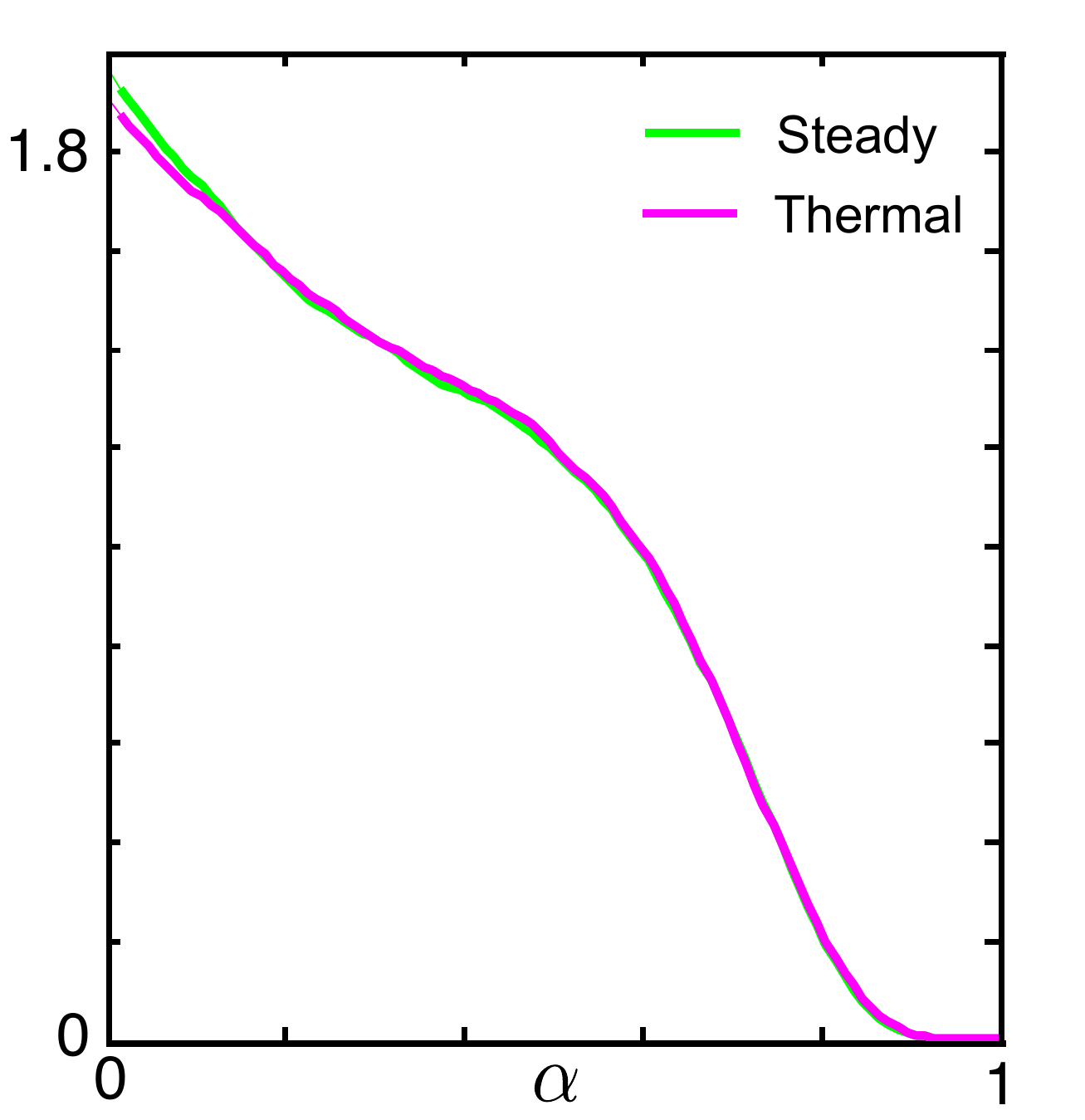}} \subfloat[]{\includegraphics[width = 1.70in]{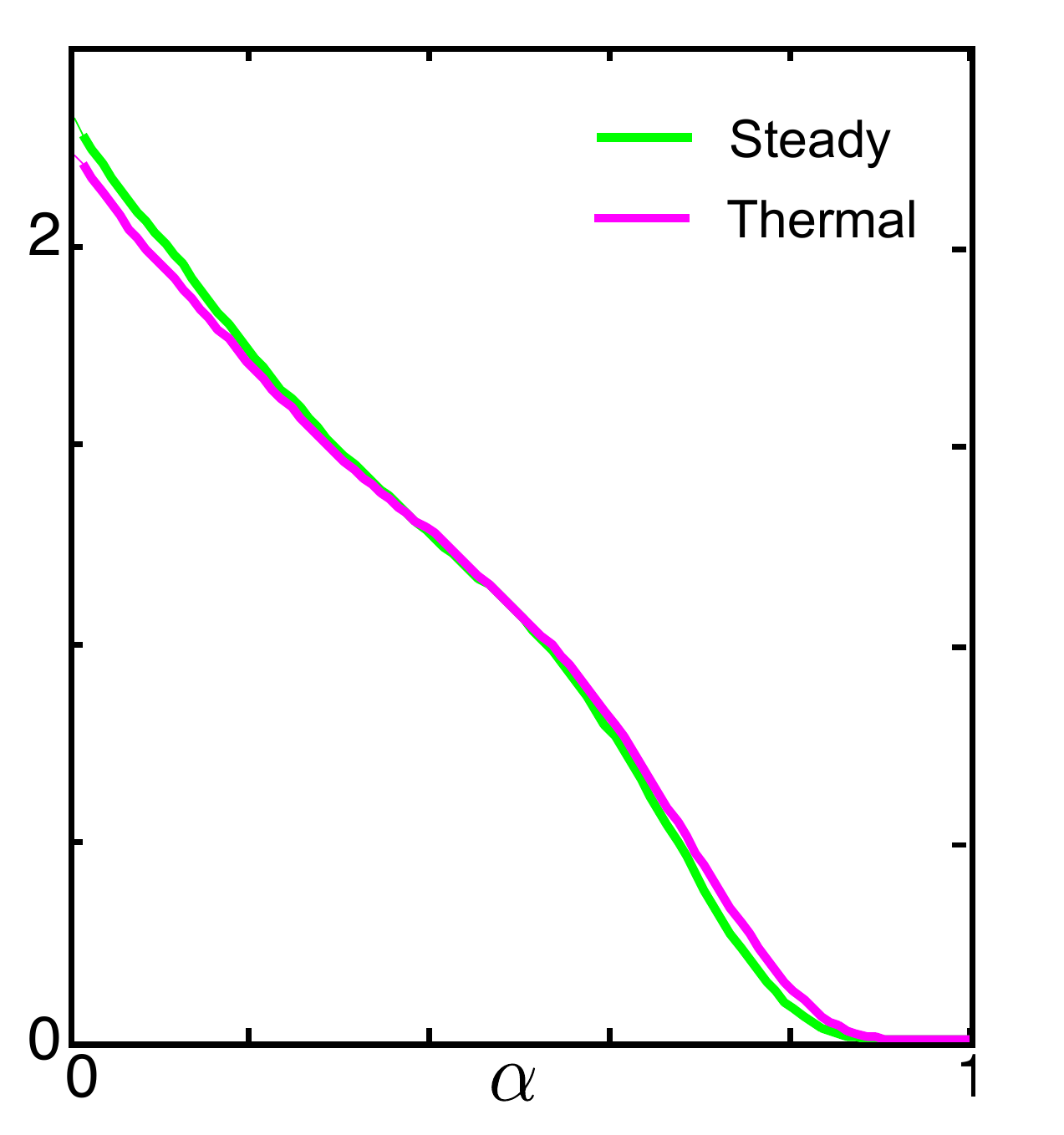}} \subfloat[]{\includegraphics[width = 1.74in]{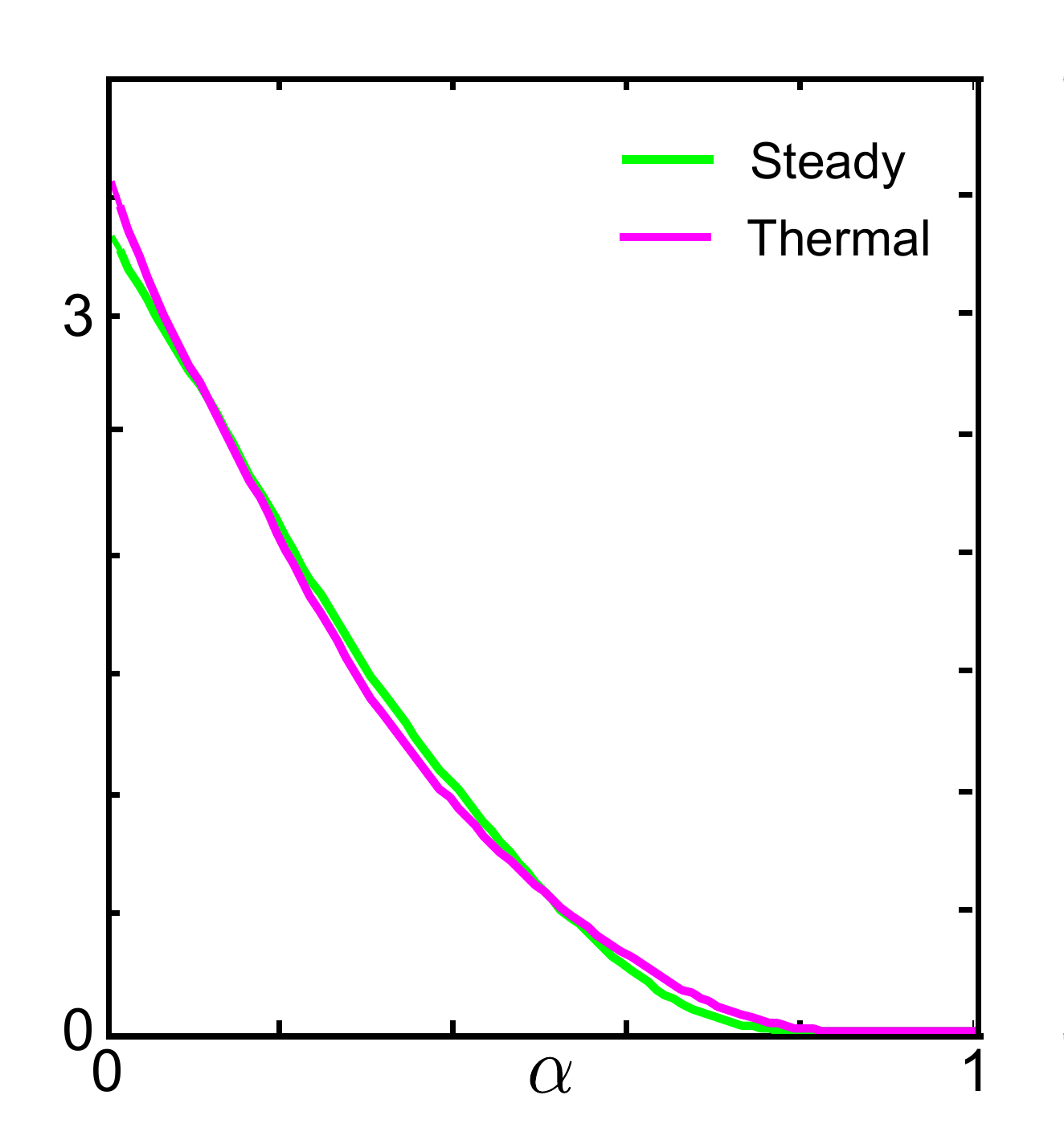}}
\caption{In comparison to plots in Fig. 1, here we use a larger integration length $l = 20 \xi$ with all other parameters unchanged. Direct comparisons between cases (a), (b) and (c) of Figs. 1 and 2 reveal that for larger integration lengths, deviations between dynamical and thermal distributions becomes smaller.}
\label{fig:equal2}
\end{figure*}
\end{center}

In Figs. (\ref{fig:equal1}) and (\ref{fig:equal2}), we compare the full distribution functions $P (\alpha, t)$ of equal time correlations, in the long time (steady state) limit, with the corresponding thermal distributions at the predicted effective temperatures (in Eq. (\ref{eq:Tans}) of the main text). For different integration lengths $l$ and velocity of splitting $v$. In all simulations, we use a large value of the Luttinger parameter $K = \rho \xi_c / 2 = 10$. We find that at large velocities ($\eta_s \sim 1$), the two distributions overlap each other near perfectly. However, at smaller velocities ($\eta_s \gg 1$), which correspond to higher effective temperatures, the distributions start to deviate from one another. This is because at higher temperatures, larger $k$ modes start to contribute to the distribution functions more significantly. The amplitude of phase fluctuations of such modes, however, deviates from the thermal result of $T/k^2$. This leads to deviations from the thermal result. These deviations are however suppressed for larger integration lengths. This is because the most important contribution to the distribution function comes from waves of momentum $k \sim 1 / l$ , which grows small for larger $l$.


\end{document}